# Advanced Accelerator Concepts

*M. Ferrario[1] and R. Assmann[1,2]*
[1]Istituto Nazionale di Fisica Nucleare – Laboratori Nazionali di Frascati, (Rome), Italy
[2]Deutsches Elektronen-Synchrotron DESY, Hamburg, Germany

**Abstract**
Recent years have seen spectacular progress in the development of innovative acceleration methods that are not based on traditional RF accelerating structures. These novel developments are at the interface of laser, plasma and accelerator physics and may potentially lead to much more compact and cost-effective accelerator facilities. While primarily focusing on the ability to accelerate charged particles with much larger gradients than traditional RF structures, these new techniques have yet to demonstrate comparable performances to RF structures in terms of both beam parameters and reproducibility. To guide the developments beyond the necessary basic R&D and concept validations, a common understanding and definition of required performance and beam parameters for an operational user facility is now needed. These innovative user facilities can include "table-top" light sources, medical accelerators, industrial accelerators or even high-energy colliders. This paper will review the most promising developments in new acceleration methods and it will present the status of on-going projects.

## 1    Introduction

Advancement in particle physics has historically been linked with the availability of particle beams of ever increasing energy or intensity. For more than three decades the collision energy in particle colliders has increased exponentially in time as described by the so-called Livingston plot. An adapted version [1] of the Livingston plot is shown in fig. 1. It includes achievements with conventional and novel accelerators and indicates the present plans beyond 2020. It is seen that particle accelerators are a remarkable success story with beam energies having increased by 5 – 8 orders of magnitude since the first RF based accelerators in the 1920s. However, it is also evident that the exponential increase of beam energy with time has levelled off in conventional accelerators since the 1980s. Limits in conventional accelerators arise from technical limitations (e.g. maximum fields in super-conducting magnets for Hadron machines, synchrotron power losses in Lepton machines, breakdown effects at metallic walls of RF cavities in Linear machines) but also practical issues like size and cost. This limitation has serious implications for future scientific use of accelerators such as TeV colliders, as it implies machines that may reach many 10's of km in length and cost in excess of 10 billion €.

To overcome such limitations a vigorous worldwide effort is on-going aimed at the development of high field superconducting magnets, of Muon colliders and of novel high-gradient acceleration techniques.

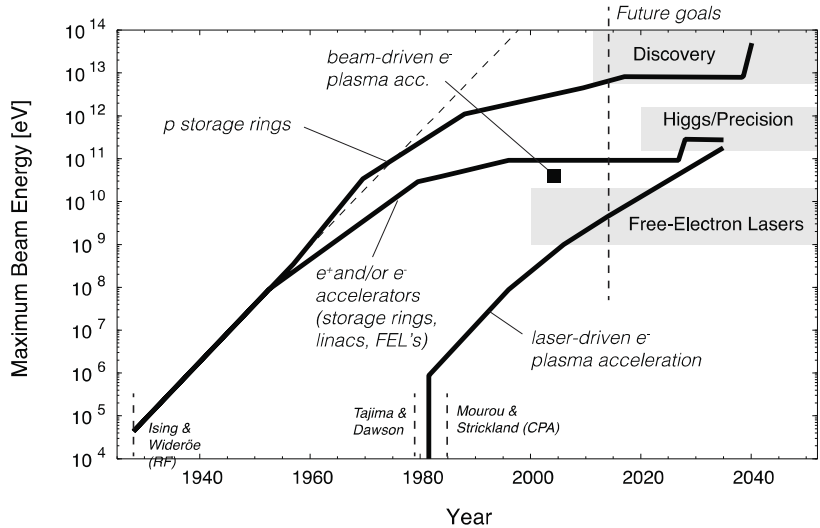

**Fig. 1: Updated** Livingston plot for accelerators [1], showing the maximum reach in beam energy versus time. Grey bands visualize accelerator applications. The left fork shows the progress in conventional accelerators from the first ideas in the 1920s. This main fork splits into two lines for electron/positron machines and for proton accelerators. A new fork of laser-driven plasma accelerators has emerged in 1980, reaching multi-GeV energies by now. Beam driven plasma acceleration results are indicated by the square point. Vertical dashed lines indicate future goals for the various technologies.

In order to afford the energy needed to excite high gradient accelerators, their operating wavelength must be reduced, from the RF to the THz spectral regime and even down to optical wavelength. This presents distinct challenges in power generation, requiring new paradigms in the excitation of accelerating waves and in the beam time structure.

In recent times a new technology emerged, based on the revolutionary proposal of Plasma accelerators by Tajima and Dawson in 1979 [2], and the invention of amplified chirped short (~fs) optical pulses (CPA) by Mourou and Strickland in mid 1980s [3], both awarded by the Nobel price for physics in 2018 for the invention of CPA techniques. Plasma-based concepts presently offer not only the high beam energies shown in fig. 1, but also the highest accelerating gradient compared to other novel acceleration techniques like high-frequency metallic RF structures or dielectric structures. To proceed towards high-energy physics (HEP) applications, however, one must demonstrate progress in beam quality and control. Advances in understanding limitations on accelerating gradient go beyond linear colliders; compact and cost-effective accelerators can be used in applications such as inverse Compton scattering gamma ray sources, free-electron lasers, medical linacs, all may benefit from larger accelerating gradients.

This paper will review the most promising developments in new high gradient acceleration methods for linear machines, namely high frequency Metallic RF structures, Dielectric structures and Plasma based accelerating modules, and it will present the status of on-going projects.

## 2    Metallic structures

Conventional normal conducting RF particle accelerators consist of metallic (copper) corrugated waveguides that can support the propagation of an electromagnetic wave with a longitudinal (accelerating) field component. Charged particles, propagating on the axis of the RF structure and synchronous with the RF wave propagation, can gain kinetic energy at the expenses of the electromagnetic stored energy. They are driven by high-power microwaves generators (Klystrons) and are currently able to achieve accelerating fields up to 120 MV/m at X-band frequency [4]. A significant progress if compared to the long-lasting S-band SLAC type structure that is typically limited to 20

MV/m. To exploit this new technology an European Design study has been recently funded, the H2020 XLS-CompactLight Project [5], which aims to design the next generation of compact X-rays Free-Electron Lasers and to investigate all the possible applications of compact high gradient X-band structures. The X-band successful progress has been achieved after an R&D effort performed mainly at SLAC [6] and CERN [4] towards the understanding of breakdown physics and improved fabrication processes. An electric breakdown event (or discharge) in fact is the main limitation in a metallic structure to achieve high gradient. The discharge event occurs from the surface melting over a macroscopic area in a high E-field region of an accelerator structure wall. A plasma forms over the molten area, bombarding the surface with an intense ion current producing consistent changes in the structure shape with crater formations in the copper surface and vacuum breakdown, thus limiting the maximum operating accelerating field.

The breakdown rate (BDR) is one of the main quantitative requirements characterizing high gradient performance of the linac. The CLIC linear collider requires RF breakdown probability to be less than 3 x 10-7/pulse/m for an accelerating gradient of 100 MV/m operating with a 180 ns long RF pulse. As a result of the R&D effort the 11.4 GHz Traveling Wave (TW) accelerating structures can today run at BDR of about 10-6/pulse/m at gradients up to 120 MV/m and ~200 ns pulse length [4]. This result scaled to a 180 ns rf long pulse is consistent with the CLIC requirement.

The pulse length $\tau_{rf}$ is one of the key parameters to achieve high gradient [7]. It has been experimentally demonstrated that the maximum achievable accelerating gradient for a fixed BDR scales like $\tau_{rf}^{-1/6}$ . Beam time structure, accelerating structure length and filling time together with the available rf peak power set the minimal acceptable RF pulse length $\tau_{rf}$. One possibility to reduce the RF peak power is to lower the pulse energy required for the same E-field in the structure by adopting structures operating at higher rf frequency, hence with a reduced volume to fill with energy. In this way the stored energy scales with RF frequency like f-3 leading to a considerable reduction of the required input peak power. On the other hand one has to face additional limitations related to fabrication tolerances and beam instability due to enhanced wake-fields effects in corrugated structures. In fact any discontinuity in the accelerating structure boundaries produces an electromagnetic disturbance induced by the beam current, called wake-field. Wake fields can have both transverse and longitudinal field components that might back interact with the beam itself causing energy spread and/or off axis deflection and eventually beam loss. Nevertheless corrugated structure boundaries are necessary to keep the phase velocity of the accelerating field synchronous with the speed of the particle beams, i. e. less than the speed of light, a condition that is impossible to satisfy in a smooth waveguide. Since the intensity of the transverse wake field is inversely proportional to the third power of the structure aperture and linearly increases with the beam offset with respect to the structure symmetry axis, higher RF frequency structures are more subject to beam instabilities than lower frequency structures. The advantage of using high frequency structures to achieve higher gradient is consequently limited by possible beam quality degradation due to wake-field effects.

Nevertheless very promising results have been achieved in this direction at SLAC at the FACET facility where rf breakdown and beam deflection studies have been performed with a 140 GHz rf structure [8], see fig. 2. Since there was not access to rf generators at that frequency, the structure has been excited by an on-axis ultra-relativistic electron beam exciting a longitudinal wake-field with a spectral component at the frequency of the accelerating mode. The maximum achieved accelerating gradient was 300 MV/m with a peak surface electric field of 1.5 GV/m and a pulse length of about 2.4 ns.

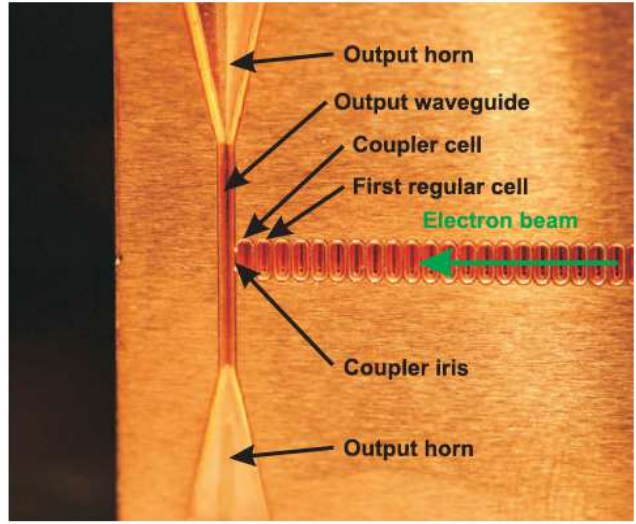

**Fig. 2:** The mm-wave traveling wave accelerating structure with detail of the output part, including coupler and output waveguides [8].

Another interesting approach to increase electric fields in copper cavities is to cool them down to temperatures below 77 K [9], where the rf surface resistance, coefficient of thermal expansion and crystal mobility decrease, while the thermal conductivity and hardness increase, all of which can contribute to reduce the BDR. Recent studies show that a X-band structure can be conditioned up to an accelerating gradient of 250 MV/m at 45 K and a breakdown rate of 2 x 10−4/pulse/m. So far for this breakdown rate, the cryogenic structure has the largest reported accelerating gradient.

A combination of short rf pulses, cryogenic and high frequency operation will probably lead to additional significant progress of high gradient rf metallic structure in the next decade. Nevertheless it is unlike to expect gradient in the range of GV/m with high frequency metallic structures because of the additional limitations related to fabrication tolerances and beam instability due to enhanced wake-fields effects in corrugated structures, as discussed above.

## 3    Dielectric structures

As discussed in the previous paragraph both the increase in operational frequency and reduction in pulse length play a role in increasing the breakdown limit. To overcome the limitations imposed by the small corrugated structures, necessary to keep the phase velocity synchronous with the particle velocity but producing enhanced wake field effects, one can consider a more simple circular metallic waveguide filled by a dielectric liner that allows matching the phase velocity of the wave to the speed of particles.

Using optical generation techniques, one can have very short THz pulses (< 100 ps) generated by picosecond lasers readily available at high average power. This approach has been adopted by the AXSIS collaboration [10] that has recently reported experimental demonstration of electron acceleration using the axial component of an optically generated 10 □J THz pulse centred at 0.45 THz propagating in a Fused Silica capillary waveguide [11]. A maximum energy gain of 7 keV has been observed in a 3 mm long structure of 400 μm inner diameter. Using more intense THz sources (∼ 10 mJ) the AXSIS collaboration foresees to reach ∼GV/m accelerating gradients in the near future.

Breakdown limits in similar dielectric structures have been investigated at FACET where Dielectric Wakefield Acceleration scheme (DWA) are under investigation. In the DWA scheme two bunches are injected in a dielectric filled waveguide so that the first high charge bunch excites a wake field that can be used to accelerate a collinear trailing lower charge bunch, see fig. 3. First measurements of the breakdown threshold in a dielectric structure, subjected to GV/m wakefields produced by short (30–330 fs) 28.5 GeV electron bunches, have been made at FACET [12].

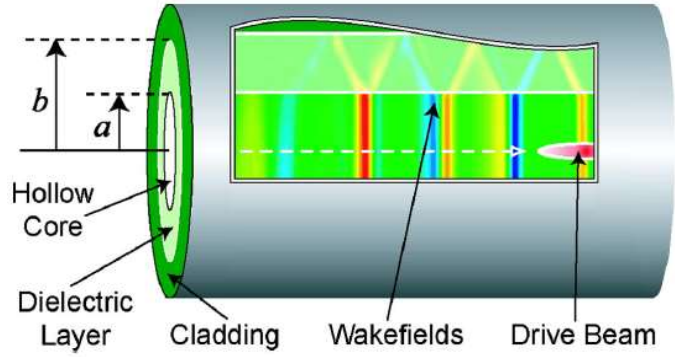

**Fig. 3**: Conceptual drawing of the dielectric wakefield accelerator (DWA) [12], a = 50 μm, b = 162 μm.

A Fused Silica capillary of 100 μm inner diameter was exposed to a range of bunch lengths, allowing surface dielectric fields up to 27 GV/m to be generated. The onset of breakdown was observed to occur when the peak electric field at the dielectric surface reaches ~13.8 GV/m. In a more recent experiment, accelerating gradient of about 1.3 GV/m have been also measured in a 15 cm long dielectric structure [13].

The use of infrared lasers to power optical-scale lithographically fabricated dielectric structures, which is referred to as dielectric laser acceleration (DLA), is another developing area in advanced acceleration techniques. This accelerating structure consists of two opposing binary gratings, separated by a vacuum gap where the electron beam travels perpendicular to the grating rulings, see fig. 4. To generate the required accelerating fields, being the dielectric material transparent to the optical radiation, a linearly polarized laser pulse is incident on the structure perpendicular to both the electron beam direction of propagation and the plane of the gratings. The structure essentially acts as a longitudinally periodic phase mask, where each grating pillar imparts a π-phase shift on the electric field. As a result, electrons launched at the correct optical phase remain phase-synchronous and experience net energy gain. In this configuration gradients beyond 250 MV/m in acceleration test of electrons in a DLA have been measured [14].

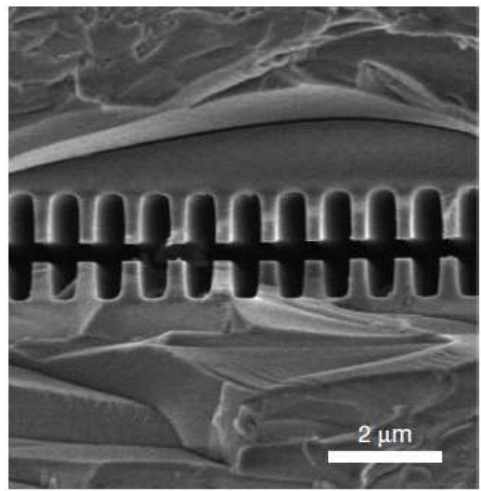

**Fig. 4**: Scanning electron microscope image of the longitudinal cross-section of a DLA structure **[14]**.

Recently the ACHIP [15] collaboration has completed the fabrication of an alternating phase focusing structure multi-staged accelerator in silicon and is started the testing [16]. The team has operated with a silicon-based tip at Stanford and with diamond coated tungsten tip at Erlangen. Both tips perform reliable and permit focusing of the electrons to sub 1 micron spot size in the accelerator structures using a lens following the emission tip. Net Acceleration and Direct Measurement of Attosecond Electron Pulses in a Silicon Dielectric Laser Accelerator are reported in [17].

More advanced design based on Photonic Band Gap structures are currently under investigation and are extensively discussed in [18].

Dielectric structures are a very promising option to achieve gradient above GV/m but more experimental data on operational breakdown rates and with accelerated beams are needed before driving realistic conclusions on the possibility to achieve high energy and high-quality beams with this technology.

## 4    Palsma structures

Plasma-based accelerators replace the metallic or dielectric walls of the accelerating structures with an ionized gas, or plasma, see Figure 5 [19].

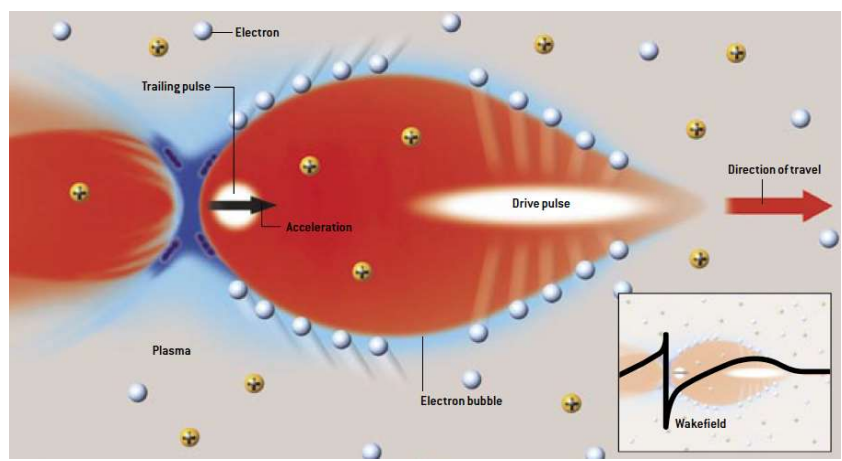

**Fig. 5**: Conceptual drawing of the plasma wakefield accelerator **[19].**

Plasma accelerator relies on a transient electron/ion charges separation excited by a driving pulse to provide the accelerating force. The drive pulse, which can be a short pulse of either a laser (LWFA [20]) or an electron beam (PWFA [21]), blows the electrons of the plasma outward, leaving behind a region of positive ion charge until the restoring force produced by the ions call back electrons towards their initial position, one plasma oscillation period later. The field produced by the ions is purely electrostatic, being the ions almost at rest in the laboratory frame (while electrons start oscillating around their initial equilibrium position) on the time scale of interest. Along the axis where the drive beam propagates, the electric field (called also plasma wake field) causes a trailing pulse of electrons injected near the rear of the bubble to undergo a very strong forward acceleration. This use of plasma to generate accelerating field allows avoiding metallic or dielectric structure damage problems due to breakdown encountered in high-gradient operation since the outer "walls" of the plasma are already "melted". For a more extensive introduction to plasma wake field accelerators see the dedicated Cas accelerator school [22]

Plasma accelerators have been tested with active length ranging from the mm to the meter scale. Accelerating gradients up to 160 GV/m have been demonstrated in experiments [23]. A number of advanced accelerator facilities are in operation or under construction in Europe (AWAKE [24], FLASH_Forward [25], SPARC_LAB [26], SINBAD [27]) and in Asia [28], complementing the two large R&D facilities that are currently spearheading advanced accelerator research in the U.S.: FACET [29] at SLAC, which is dedicated to PWFA studies and BELLA [30] at LBNL, which leads the field of LWFA.

Three fundamental milestones have been recently achieved:

- At BELLA, multi-GeV electron beams with energy up to 7.8 GeV, 6% rms energy spread, 5 pC charge, and 0.2 mrad rms divergence have been produced from a 20 cm long laser-heated capillary discharge waveguide with a plasma density near to 3 x$10^{17}$ cm$^{-3}$, powered by 850 TW laser pulses [31].
- At FACET (PWFA), acceleration of about 74 pC of charge contained in the core of the trailing bunch in an accelerating gradient of about 4.4 GV/m has been demonstrated. The core electrons gained about 1.6 GeV of energy per particle, with a final energy spread as low as 0.7%, and an energy transfer efficiency from the wake to the bunch that can exceeds 50% [32].
- More recently at FACET, using a nonlinear plasma wake driven by a single positron bunch, a substantial number of positrons has been accelerated and guided over a meter-scale plasma, in a unique and unexpected new collective regime [33].

To proceed towards high-energy physics (HEP) applications, however, one must demonstrate progress in beam quality and control. Indeed, for any variant of plasma wakefield accelerator to be practical as a Linear Collider, a range feasibility and practicality issues must be resolved in the context of an integrated system test. Plasma accelerators, like standard accelerator modules, must be capable of being staged in a series of segments [34]. Both PWFA and LWFA approaches must demonstrate simultaneous electron and positron acceleration with stable focusing in plasma and in transport lines [35, 36, 37] with performance consistent with preserving electron and positron beam quality [38, 39]. Both must demonstrate timing, pointing, and focusing control that fulfil the demands of high luminosity operation required by a lepton collider. Finally, both must demonstrate that single- and multi-bunch plasma instabilities can be overcome with operation at the tens of kHz repetition rate required for high luminosity. Beyond the feasibility issues are practical questions related to overall cost, efficiency, and reliability.

To this end new initiatives are now arising, such as the EUPRAXIA program [40, 41], which seek, through intermediate application goals, to push plasma accelerators from an exciting concept to a mature approach. It is in fact widely accepted by the international scientific community that a fundamental milestone towards the realization of a plasma driven future Linear Collider will be the integration of a high gradient accelerating plasma modules in a short wavelength Free Electron Laser (FEL) user facility [42]. The capability of producing the required high quality beams [43, 44, 45] and the operational reliability of the plasma accelerator modules will be certainly certified when such an advanced radiation source will be able to drive external user experiments. The realization of such a new generation light source thus serves as a required stepping stone for HEP energy applications and it is a promising new tool for photon science in its own right. To this end the Italian accelerator community, in the framework of the European initiative EuPRAXIA [46,47], is now giving important contributions in this direction with the recently established project named EuPRAXIA@SPARC_LAB [48, 49] under design at the INFN Laboratori Nazionali di Frascati (LNF).

## 5    Conclusions

Accelerator-based High Energy Physics will at some point become practically limited by the size and cost of the proposed e$^+$e$^-$ colliders for the energy frontier [50]. Novel Acceleration Techniques and Plasma-based, high gradient accelerators open the realistic vision of very compact accelerators for scientific, commercial and medical applications. The R&D now concentrates on beam quality, stability, staging and continuous operation. These are necessary steps towards various technological applications. To this end pilot user facilities are a planned in the near future. A major milestone is an operational, 1 GeV compact accelerator. This unit could become a stage in a high-energy accelerator.

The progress in advanced accelerators will benefit also from strong synergy with general advances in technology, for example in the laser [51] and/or high gradient RF structures industry.

To accomplish the final dream of a compact and cost effective accelerator a parallel effort should be addressed also on the development of compact devices such as beam focusing and manipulations elements, diagnostics components and eventually short period undulators [52].

Advanced accelerator science and technology is certainly an exciting and fast growing field of research [53] that stimulates the interest of an increasing number of young scientists and research institutes. We are confident that in near future a large number of our dreams will come true.

## Acknowledgements

Greatly appreciate suggestions and discussions with: E. Gschwendtner, W. Leemans, D. Alesini, B. Spataro, A. Mostacci, A. Cianchi, E. Chiadroni, C. Vaccarezza and the EuPRAXIA, XLS, BELLA, FLASH_Foward, SINBAD, AWAKE, AXSIS and EuPRAXIA@SPARC_LAB collaborations.

This work was partially supported by the European Union's Horizon 2020 Research and Innovation programme under grant agreement No. 653782 (EuPRAXIA) and under grant agreement No 777431 (XLS).